# On the Possible Common Nature of Double Extensive Air Showers and Aligned Events.

V. I. Yakovlev

*Lebedev Institute for Physics, Russian Academy of Sciences, Moscow, Russia*

**Abstract** Double Extensive Air Showers and aligned events were discovered at energies E $\gtrsim 10^{16}$ eV over fourth century back. But up to now there is no sufficiently identical explanation of their nature. In this paper it is expected that both types of events are the result of breakup of the string formed in the collisions of super high energy particles.

## Double Extensive Air Showers

Double extensive air showers (D-EAS) were discovered 26 years ago in experiment of Yoshida et al [1]. The authors of [1] arrived at the conclusion that these events were initiated by a heavy particle of mass equal to a few tens of GeV units and lifetime $5 \times 10^{-7}$–$2 \times 10^{-6}$ s. Many years of D-EAS investigations [1-3] did not result in a conclusive explanation for observed results. The conclusion [1] that delayed showers are produced by long-lived heavy particles is based on rather general ideas. The estimates obtained in [1] for the parameters of such long-lived heavy particles are questionable. Other authors [2-3] refrain from making such estimations.

In recent paper of Yakovlev et al. [4-5] there were detected 98 D-EAS from total 2117. If one assumes that a delayed shower is a result of a random coincidence with a small local shower (or with any other random reasons) then the probability for the detection of such a shower by one detector can be found from results: 98 : 2117 = 0.04629. The expected number of such random showers detected simultaneously by two detectors would then be $(0.04629)^2$ : 2117 = 4.536, as contrasted against 14 recorded (4.4$\sigma$). The expected number of events recorded by three detectors simultaneously is 0.21 versus six recorded events. Finally, the expected number of events recorded by all four detectors is 0.0097, as against three recorded events (30$\sigma$). Examples of D-EAS recorded in 2 and 4 scintillator detectors are shown in fig.1-2.

In papers [4-5] it was shown that the mass of assumed heavy particle initiated delayed EAS must be above ~$10^5$ GeV. This value was evaluated on the base of 2 safely measured values: delayed shower velocity (using delay time equal 109±9 ns and extent of shower way in the atmosphere) and delayed shower energy $E$ (using shower size $N$~$10^7$ par-ticles). The estimated value of mass is enormous, therefore, authors of [4-5] put forth two assumptions on the nature of delayed giant showers.

According to the first assumption, these showers may be initiated by a large number of low-energy hadrons produced, for example, upon the release of quark–gluon plasma. According to the second assumption, the "delayed" shower is produced by ordinary hadrons which move at a speed equal to the speed of light, while the "leading" shower is produced by a tachyon originating from the primary interaction and having a speed higher than the speed of light. The tachyon undergoes acceleration losing energy, and the shower initiated by it comes ahead of the shower initiated by ordinary hadrons. According to [6], a charged tachyon loses the entire amount of its energy by Vavilov–Cherenkov radiation. According to the classic study of Tamm and Frank [7], the energy loss by radiation per unit length is

$$dE/ds = -4\pi^2 Z^2 e^2 (\int (1 - c^2/v^2 n^2) v dv)/c^2,$$

where $Ze$ is the charge of the moving particle ($e$ is the electron charge), $n$ is the coefficient of refraction of the medium being considered, and $v$ is the frequency of emitted radiation.

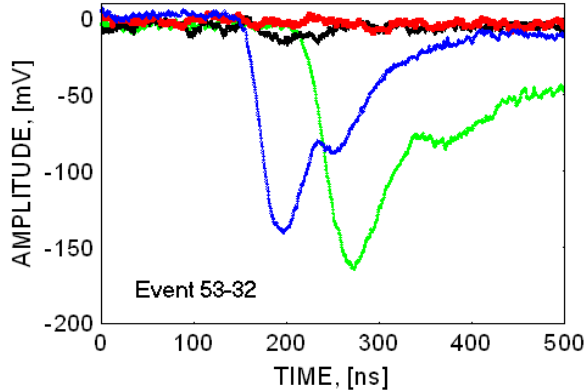 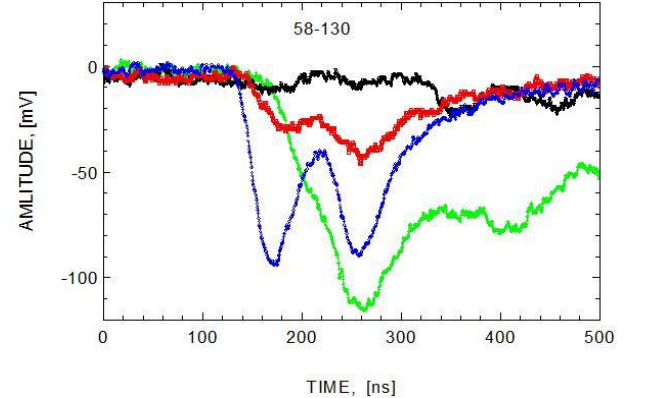

*Fig.1. D-EAS in 2 scintillator detectors*   *Fig.2. D-EAS in 4 scintillator detectors*

It follows that, as the velocity $v$ of the particle grows, the rate of its energy loss becomes higher, and this leads to a very fast formation of a leading shower. In this case, it is natural to assume [5] that a shower initiated by charged tachyon does not contain muons. A discovery of such muon-free leading showers would be strong evidence in favor of the hypothesis that it is tachyons that initiate leading showers.

It is necessary to emphasize that D-EAS were not observed at energies $E < 10^{16}$ eV.

## Aligned Events

Aligned events were firstly founded (1986) in large scale experiment with X-ray emulsion chambers (XREC) of lead and carbon type performed by the PAMIR collaboration (4400 m a.s.l.) [8].

The alignment is the location of so-called energy distinguished cores (EDC) or high energy particles along a straight line on X-ray film in gamma ray–hadron (γ-h) super families (groups of particles with energy E ≥ 4 TeV) in EAS cores with energies $\sum E_\gamma \geq 700$ TeV.

Later such aligned events were founded as at mountains [9], so in the upper atmosphere [10-11]. It should be noted that stratospheric experiments observe the alignment of particles, while mountain experiments investigate the alignment of EDCs originated by these particles. Examples of aligned events are shown in fig. 3-4. To analyze the alignment, the parameter

$$\lambda_N = \sum_{i \neq j \neq k}^{N} cos2\varphi^k_{i,j} / [N(N-1)(N-2)]$$

was introduced [12], where $\varphi^k_{i,j}$ is the angle between vectors issuing from $k$-th to $i$-th and $j$-th points. Value of $\lambda_N$ decreases from 1 (for $N$ points disposed along single straight line) to $-1/(N-1)$ (in isotropic case). Events are referred to as aligned, if inequality $\lambda_N \geq \lambda_{fix}$ is valid for their $N$ most energetic objects. Most often, $N = 4$ and $\lambda_{fix} = 0,8$. Actually, events with $N = 4$ were chosen only because of too high statistical background for $N \leq 3$ and rather poor statistics for $N \geq 5$. The threshold-like behavior of the effect has been observed: no alignment at γ-family energies $\sum E\gamma \leq 100$ TeV, then very slow increase within energy range $100$ TeV $\leq \sum E\gamma \leq 500$ TeV and rather rapid increase from $\sum E\gamma \geq 500$ TeV to manifest itself finally in 20-40% of total number of events (Table 1, fig.7).

It was shown by Mukhamedshin [13] that the phenomenon cannot be explained with cascade development fluctuations and external fields (geomagnetic field, lightning) within the framework of present-day versions of quark-gluon string models as well as without assuming the manifestation of a process in hadron interactions characterized with large transverse momentum in the Lab frame of reference at energies $s^{1/2} > 4$ TeV. There had been used different approaches to explain the nature of aligned events [14-17].

The first attempt of theoretical consideration of the above alignment phenomenon has been made by Halzen and Morris [14], whose approach was based on the assumption that semihard gluon jets is a feature of all events at energies above $10^4$ TeV. The model [14] seems to be not serious as the observable alignment of the most energetic EDCs cannot be originated by low-energy gluon-jet particles.

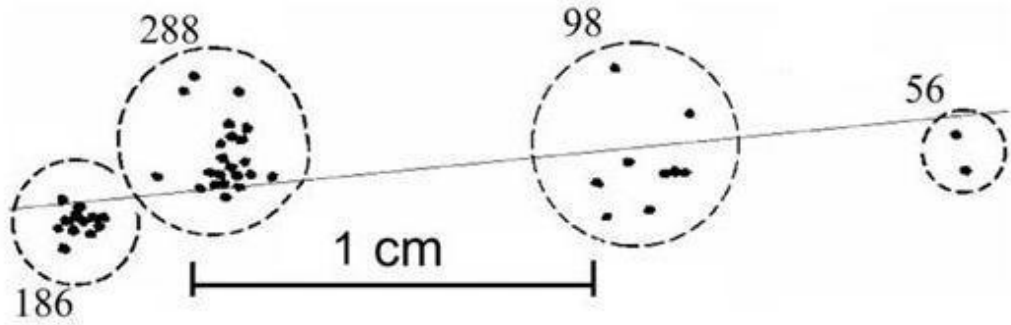

*Fig. 3. Four γ-cluster event. Digitals mean energy in TeV.*

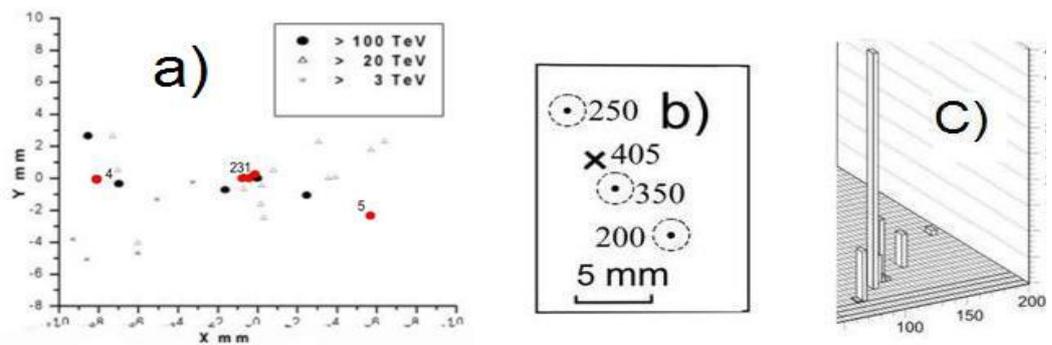

*Fig.4. a) STRANA event (balloon); b) Pb-6 XCREC; c) JF2af2 event(Concord)*

Royzen proposed a model of semi-hard double diffraction inelastic dissociation (SHDID) [15], which assumes the alignment to be a result of a tension of a quark - gluon string in the diffraction cluster between a semi-hardly scattered constituent quark and other spectator quarks of the projectile hadron and its following rupture (fig.5).

However, the present-day version seems to be incapable for providing a sufficiently large SHDID cross section to explain experimental data. As it is shown in [15] the fraction of SHDID is expected to be $\sigma_{DD}/\sigma_t \sim 0.04$; 0.07; and 0.10 at $s = 10^5$; $10^6$ and $10^7$ GeV$^2$ respectively.

It can be several times less or larger, since the above estimate is rather rough, but its smooth logarithmic threshold-like energy increase is independent of the choice of parameters.

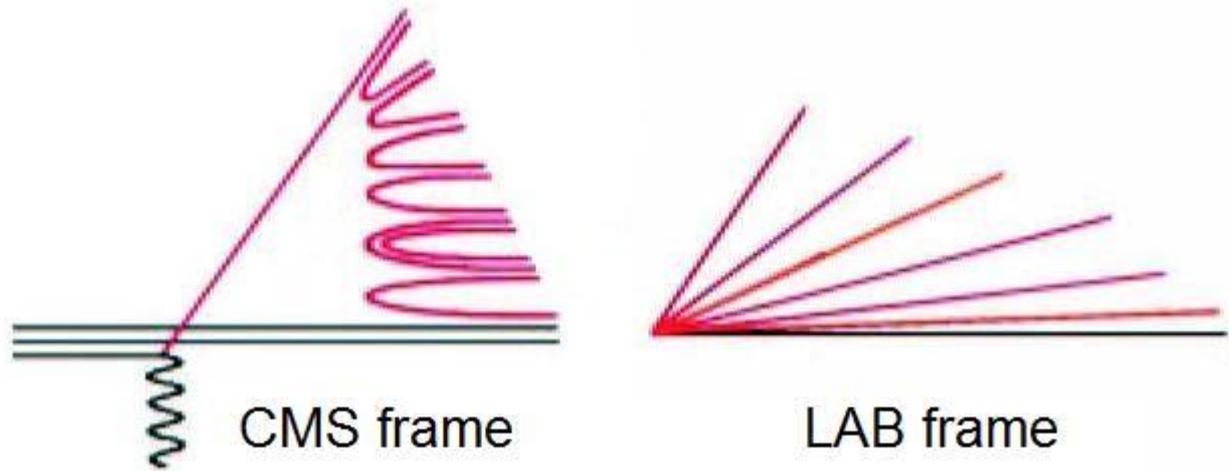

*Fig.5. Semi-hard double inelastic diffraction.*

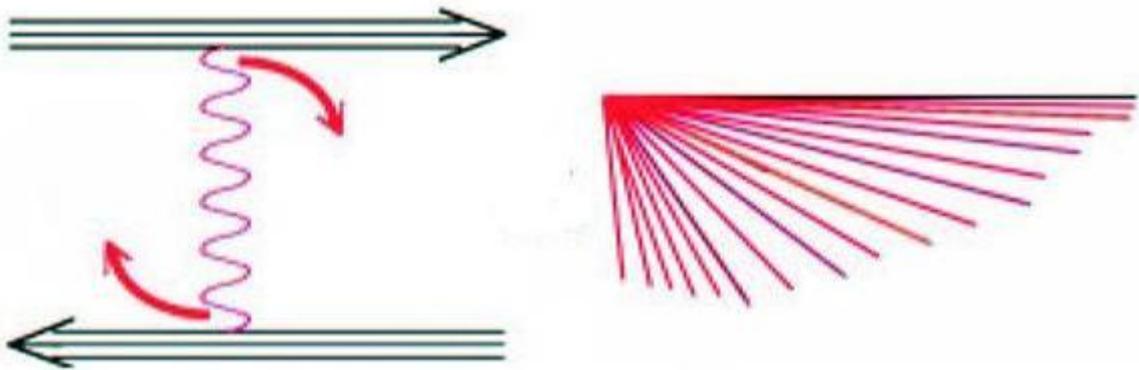

*Fig.6. Fast rotating quark-gluon string*

The rotating nuclei fragmentation hypothesis of Erlykin and Wolfendale [16] is naturally connected with the postulated increase of the fraction of heavy nuclei in the primary cosmic ray flux around $10^{16}$ eV (up to 50%). However, due to the existence of the few TeV threshold in the X-ray film technique, the experiments noticeably favour primaries of higher energy per nucleon (for the same energy per particle) and the existence of heavy nuclei alone is not enough to account for the 30% alignment that is needed. The problem was discussed in detail in Ref. [16]. The mechanism of delayed fragmentation of fast rotating nuclei needs to be established theoretically as well as the possibility of a reduction of its cross section for interaction with air nuclei. Additionally, the observation that aligned events are more abundant in the vertical component, supports the concept that the origin of the phenomenon is a deeply penetrating cosmic ray particle (most likely a proton) because alignment blurring in atmosphere.

Wibig [17] postulates no "new physics" in discussed phenomenon. The conservation of the angular momentum in the creation of fast rotating strings (fig.6) leads to its co-planar decay. The problem of quantitative description of the hadronization of such object needs detailed knowledge of the nature of the string – chain, fireball or jet. Each of these words has its individual connotation and it has not yet been decided which (if any) describes the high energy particle production process.

A number of other models exist for aligned phenomenon explanation.

Table 1.

| **FRACTION OF ALIGNED EVENTS** |
|---|
| **PAMIR** ($\sum E_\gamma \geq 700$ TeV, $\lambda_4 \geq 0,8$) |
| 0,43±0,13 in Lead XREC (6 from 14, expected 1) |
| 0,22±0,05 in Carbon XREC (13 from 59, expected 2,1) |
| Expected background 0,06 |
| **KANBALA** ($\sum E_\gamma \geq 500$ TeV, $\lambda_3 \geq 0,8$) |
| 0,5±0,2 in Ferrum XREC (3 from 6, expected 1,2) |
| Expected background 0,21 |
| **TWO STRATOSPHERIC EVENTS** at $\sum E_\gamma \geq 1000$ TeV |
| $\lambda_4(\gamma) = 0,998$ (event JF2af2, Concord) |
| $\lambda_4(h) = 0,99$ (event STRANA, balloon) |

To estimate the influence of large transverse-momentum processes, a coplanar particle generation model (CPGM) was used by Mukhamedshin [13] with the following features: (a) the multiplicity is $<n> \approx 10$; (b) the mean transverse momentum of five most energetic particles in the coplanarity plane is $<p_t^{copl}> = 2.34$ GeV/c; (c) the transverse momentum, normal to the coplanarity plane, is $<p_t> = 0.4$ GeV/c; (d) the CPG interaction occurs in each of proton initiated cascades at $E_0 \geq E_{0thr}^{copl} = 8$ PeV. Comparison of model with Pamir data are shown in fig.7. Different variants of CPG models were considered in FANSY-code model by Mukhamedshin [18].

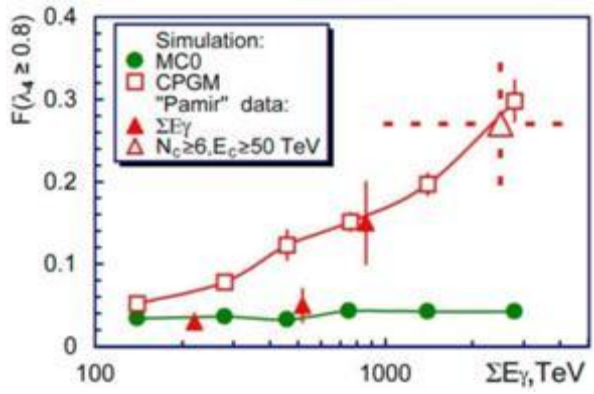 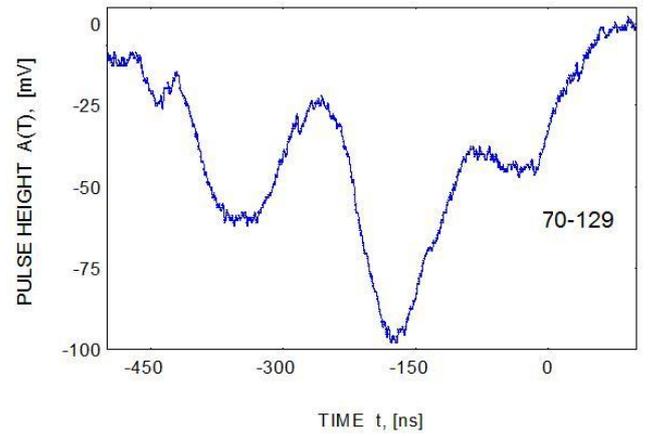

*Fig.7. Alignment dependence on $\sum E_\gamma$.*   *Fig.8. Triple EAS in Cherenkov light.*

## On the Possible Nature of D-EAS and Aligned Events

As both phenomena demonstrate the threshold-like behavior at energies $s^{1/2} > 4$ TeV it is possible to assume with high degree of probability that they have the uniform general nature. Presence of two features of a phenomenon: the spatial (alignment-coplanarity) and time delay can facilitate finding-out of true nature of the phenomenon. At absence of the adequate theory explaining these phenomena it is necessary to use some phenomenological (or heuristic) approach for their description.[1]

At present time the theory, pretending on the united theory of the whole is the theory of strings [19-20]. According to boson theory of strings tachyon inserts in its frames. This theory expects that boson strings have space-temporary dimension, equal 26. However this variant of theory of strings does not describe fermions.

Theories, describing simultaneously, both bosons and fermions, are called theories of superstrings. There are 5 such theories at present. Usage duality principle allows to find out that all 5 theories can be possibly agreed with each other. These five theories are different limiting cases of united fundamental theory, which got name M-theory. All 5 theories have dimensionality, equal to 10, and exclude tachyon existence outside of the vacuum.

For this reason we turned to the attempt of other explanation of D-EAS nature, so as aligned events.

---

[1] Heuristic ways of the decision of problems name the ways, allowing to minimize search of the possible decisions, frequently based on intuition.

When we "dislodge" one quark or gluon, it "wants" to fly off, but cannot, as far as it must for itself pull "string", which contains the bag of virtual particles which is bigger, as greater the energy of a projectile particle. This "string" can be pulled between the fly object, which we observe, and remainder of the proton. After string breakup both its parts get pulses of opposite sign, and some kind of sling effect arises. This is the reason why each of D-EAS pairs moves with slightly different velocity. "Delayed" part of string initiates "delayed" shower. The "leading" part of string which has received an additional impulse, starts to extend even faster, and this extend accelerates with each next particle materialization. Thus "leading" part of string produces aligned particles and later "leading" shower in the atmosphere. Time delay between leading and delayed showers must depend on the value of strain between two parts of string. Possibly, that value of delay can be used for the estimation of string strain power and some other properties of string.

If we consider that leading shower flies at the speed of the light, then delayed for 100 ns shower must fly (refer to [5]) at the speed of $2.99 \times 10^{10}$ cm/s. Thereby, absolute difference of velocities of leading and delayed showers (at the value of delay ~ 100 ns) forms $\Delta v = 10^8$ cm/s = $10^6$ m/s. Thence possible estimation of a total value of additional pulse $\Delta p$ got at the string breakup (for the given concrete event) $\Delta p = m\Delta v /\gamma$ for free particle with the mass m.

As it was shown in [4-5] the mass of particle initiated delayed for 100 ns shower with $N\sim 10^7$ particles should be above $10^5$ GeV. For such particle (or string bag) $\Delta p = (10^5 \times 10^6)/15$ GeV·m/s = $6.6 \times 10^9$ GeV·m/s.

Because of too high statistical background by $N < 3$ (where $N$ is the number of EDC) alignment phenomenon investigates at $N \geq 3$ most energetic particles. Such aligned events appear after materialization of particles from leading string bag.

At investigation of D-EAS with Tien-Shan array of Vavilov-Cherenkov radiation triple EAS's were detected [21] when observing under zenith angle $\theta=70^o$ to vertical. Example of triple EAS is shown in fig.8.

## Search of D-EAS and ALIGNED events at LHC

As it has been shown above at development of D-EAS in atmosphere the time delay for 109 nanoseconds collects between two EAS's. If to consider, that both EAS were started at height of 20 km above the observation level (23,33 km above sea level, x=35 g/sm$^2$) the delay accumulated for way to 1 m, will make $109/2·10^4 = 0.00545$ ns/m. Thus, at registration in detector CASTOR (14.38 m from collision point) delay will make 0,078 nanoseconds, and at registration in detector ZDC (140 m from interaction point) delay will make 0.763 nanoseconds. For detector placed at distance 1000 m from collision point the delay time between two "showers" will reach 5.45 ns, enough for reliable resolving.

Extremely high value of alignment about 1 (for almost all points disposed along single straight line!) in stratospheric event "Strana" and in upper atmosphere event "Concord" permits to expect that such events would be easy discovered at LHC.

## Conclusion

D-EAS and aligned events appear at the same energy $E \gtrsim 10^{16}$ eV. They can arise as the result of breakup of string which forms at interaction of super high energy particles. Value of the pulse at which string breakup can occur was estimated: $\Delta \boldsymbol{p} = 6.6 \times 10^9$ GeV·m/s.

No model existing at present time can explain extremely high alignment in stratospheric events "Strana" and "Concord". Such events would be easy discovered at LHC.

## Acknowledgment

Author expresses his thanks to R.A. Mukhamedshin for valuable comments on aligned events.

## References

1. M. Yoshida, Y. Toyoda, and T. Maeda, *J. Phys. Soc. Jpn.* **53**, 1983 (1984).
2. O. V. Vedeneev, Yu. A. Nechin, Yu. A. Fomin, and G. B. Khristiansen, *VANiT, Ser. Tekh.Fiz. Eksperim.,* No. **3(29),** 47 (1986).
3. M. Ambrosio, C. Aramo, L. Colesanti, and A. D. Erlykin, *Nucl. Phys. B* Proc. Suppl. **52B**, 234 (1997).
4. V. I. Yakovlev, M. I. Vil'danova, and N. G. Vil'danov,

*Pis'ma Zh. Eksp. Teor. Fiz.* **85**, 111 (2007) [*JETP Lett.* **85**, 101 (2007)].

5. V. I. Yakovlev, M. I. Vildanova, N. G. Vildanov, and A. V. Stepanov, *Physics of Atomic Nuclei*, (2010), Vol. **73**, No. 5, pp. 785–790.
6. T. Alvager and M. N. Kreisler, Phys. Rev. **171**, 1357 (1968).
7. I.E. Tamm and I.M. Frank, Dokl. Akad. Nauk SSSR **14**, 109 (1937).
8. Pamir collaboration, *Proc. of IV-th ISVHECRI*, Beijin 1986.
9. L. Xue et al., *Proc. of 26th ICRC, Salt Lake City*, **1** (1999) 127.
10. A.V. Apanasenko et al., *Proc. of 17$^{th}$ ICRC, Plovdiv* **7** (1977) 220.
11. J.N. Capdevielle, *J. Phys.* **G 14** (1988) 503.
12. A.S. Borisov et al. *Proc. of 8th ISVHECRI, Tokyo (1994) 49.*
13. R.A. Mukhamedshin, *JHEP 05 (2005) 049.*
14. F. Halzen and D.A. Morris, Phys. Rev. **D 42** (1990).
15. I.I. Royzen, *Mod.Phys.Lett.* **A 9** (1994) 3517.
16. A.D. Erlykin and A.W. Wolfendale, Nucl. Phys. B (Proc. Suppl.) **75A**, 209 (1999).
17. T. Wibig, arXiv: hep-ph/0003230v1 23 Mar 2000.
18. R.A. Mukhamedshin, Eur. Phys. J. C (2009) **60**, 345-358;
19. M. Kaku, *Introduction to Superstrings*, Springer-Verlag.
20. B. Greene, *The elegant Universe*, Vintage Books, New York.
21. R.U. Beisembaev, Yu. N. Vavilov, M. I. Vildanova, N. G. Vildanov, V. V. Zhukov, A. V. Kruglov, R. A. Nam, V. P. Pavlyuchenko, V.A. Ryabov, A. V. Stepanov, Zh. S. Takibaev, V. I. Yakovlev. *Izvestiya of Russian Academy of Sciences*, (2011, **V.75,** №3, 385-387) in print.